%% file: main.tex
\documentclass[journal,letter,10pt,twocolumn]{IEEEtran}
\usepackage{amsmath,graphicx}
\usepackage{bbm}
\usepackage{amsmath, amssymb, amscd, amsthm, amsfonts}
\usepackage{subfiles}
\usepackage{color}
\usepackage{booktabs}
\usepackage{multirow}

\usepackage[caption=false,font=footnotesize]{subfig}
\pdfoutput=1

\usepackage[backend=biber, style=ieee, dashed=false,isbn=false,doi=false,url = false,bibencoding=utf8]{biblatex}
\title{Symbol Detection Using an Integrate-and-Fire\\ Time Encoding Receiver}

\author{
\IEEEauthorblockA{Neil Irwin Bernardo}\\
\IEEEauthorblockA{\textit{Electrical and Electronics Engineering Institute} \\
\textit{University of the Philippines Diliman}\\
Quezon City, Philippines\\
neil.bernardo@eee.upd.edu.ph}
\thanks{The author acknowledges the Office of the Chancellor of the University of the Philippines Diliman, through the Office of the Vice Chancellor for Research and Development, for funding support through the PhD Incentive Award Grant 252509 YEAR 1.}
}
\begin{document}
\pagenumbering{gobble}
%\ninept
%
\maketitle
\begin{abstract}
Event-driven sampling is a promising alternative to uniform sampling methods, particularly for systems constrained by power and hardware cost. A notable example of this sampling approach is the integrate-and-fire time encoding machine (IF-TEM), which encodes an analog signal into a sequence of time stamps by generating an event each time the integral of the input signal reaches a fixed threshold. In this paper, we propose a receiver architecture that estimates the sequence of transmitted symbols directly from the encoded time stamps, called \emph{time encodings}, produced by the IF-TEM sampler on the received signal. We show that waveform reconstruction from time encodings is not necessary for symbol detection. We develop an analytical approximation for the symbol error probability (SEP) of the proposed IF-TEM-based receiver and show that it closely matches the SEP results obtained through Monte Carlo simulations. Additionally, we demonstrate that narrowing the 3 dB bandwidth of the transmit pulse shaping filter degrades the proposed IF-TEM receiver’s performance, highlighting a trade-off between spectral efficiency and error resilience.
\end{abstract}
\section{Introduction}
\label{section:intro}

\subfile{Sections/introduction.tex}

\section{System Model and IF-TEM Receiver Structure}
\label{section:sys_model}

\subfile{Sections/problem_setup.tex}

\section{Symbol Error Probability Analysis of the Proposed IF-TEM Receiver}
\label{section:error_analysis}
 \subfile{Sections/error_analysis.tex}

\section{Results and Discussion}
\label{section:numerical_results}

\subfile{Sections/numerical_result.tex}

\section{Summary and Future Work}
\label{section:conclussion}
\subfile{Sections/conclusion.tex}

\renewcommand*{\bibfont}{\footnotesize}
\begingroup
\footnotesize  % or \small, \scriptsize, etc.
\printbibliography
\endgroup
%\bibliographystyle{IEEEtran}
%\bibliography{refs}

% -------------------------------------------------------------------------
% \begin{figure}[htb]

% \begin{minipage}[b]{1.0\linewidth}
%   \centering
%   \centerline{\includegraphics[width=8.5cm]{image1}}
% %  \vspace{2.0cm}
%   \centerline{(a) Result 1}\medskip
% \end{minipage}
% %
% \begin{minipage}[b]{.48\linewidth}
%   \centering
%   \centerline{\includegraphics[width=4.0cm]{image3}}
% %  \vspace{1.5cm}
%   \centerline{(b) Results 3}\medskip
% \end{minipage}
% \hfill
% \begin{minipage}[b]{0.48\linewidth}
%   \centering
%   \centerline{\includegraphics[width=4.0cm]{image4}}
% %  \vspace{1.5cm}
%   \centerline{(c) Result 4}\medskip
% \end{minipage}
% %
% \caption{Example of placing a figure with experimental results.}
% \label{fig:res}
% %
% \end{figure}

%\vfill
%\pagebreak

%\vfill\pagebreak


\end{document}

%% file: Sections/introduction.tex
Traditional signal acquisition systems rely heavily on uniform sampling methods, most notably those guided by the Shannon–Nyquist Sampling Theorem \cite{Shannon:1949}. While this framework ensures perfect reconstruction of bandlimited signals under ideal conditions, uniform sampling methods often necessitate high sampling rate and precise clocking, which can be inefficient or impractical for power-constrained and low-complexity devices \cite{Koscielnik:2007,Tsividis:2010}. As the demand for energy-efficient, scalable, and asynchronous signal acquisition grows, alternative sampling paradigms have gained traction.

Event-driven sampling, wherein samples are acquired based on signal activity rather than a fixed clock, has been considered as a promising alternative to clock-driven uniform sampling schemes \cite{miskowicz2015event}. One biologically inspired realization of this principle is the Time Encoding Machine (TEM), particularly the Integrate-and-Fire Time Encoding Machine (IF-TEM), which generates a time stamp each time the integral of the input signal reaches a predetermined threshold \cite{Lazar:2004}. This asynchronous mechanism mimics neural spike generation \cite{Tuckwell_1988} and enables sampling systems that operate with lower energy consumption and more relaxed hardware requirements compared to conventional uniform sampling methods. 

Early works on TEM focused on the development of recovery algorithms and the establishment of their recovery guarantees for reconstructing band-limited (BL) signals from time encodings \cite{Lazar:2004,Lazar:2004b, Thao:2021}. Recovery algorithms for TEM have also been developed for finite-rate-of-innovation (FRI) signals, signals in shift-invariant (SI) spaces, and sparse signals \cite{Gontier:2014,Alexandru:2021,Kamath:2022,Naaman:2022,Florescu:2023,Kamath:2023}. In addition to theoretical advancements, practical implementations of TEMs have been explored in neuromorphic hardware, data converters, image/video processing, and biomedical engineering, where power efficiency and asynchronous operation are critical \cite{Costas-Santos:2007,Lazar:2011,Kong:2012,Lazar:2014, Gutierrez-Galan:2022,Fu:2024,Naaman:2024,Naaman:2024b,Arora:2025}.

Despite the promising benefits of TEM, the integration of IF-TEMs into communication receiver architectures and the characterization of the communication performance of IF-TEM-based receivers remain relatively unexplored. Conventional digital communication receivers typically assume uniformly-sampled baseband signals and rely on digital signal processing (DSP) pipelines to recover the transmitted information. Incorporating IF-TEMs into the receiver front-end disrupts this convention since the output of the sampler is no longer a fixed-rate sequence but a set of irregularly-spaced time stamps. This poses new challenges in symbol detection, equalization, and synchronization. Recent work has demonstrated a fully neuromorphic, repetition-coded BPSK receiver that jointly performs detection and decoding from event-based samples \cite{Katsaros:2025}. However, this approach has not yet been extended to multi-level signaling schemes or to pulse shapes beyond rectangular pulse.

%posing new challenges in symbol detection, equalization, and synchronization, as the output of the sampler is no longer a fixed-rate sequence but a set of irregularly-spaced time stamps.

In this work, we explore the feasibility of performing symbol sequence detection directly from time encodings of the received continuous-time pulse amplitude modulation (PAM) waveform, without reconstructing the underlying continuous-time waveform. Specifically, we propose an IF-TEM-based receiver that detects symbols based on the number of firing events within each symbol interval, and mitigates inter-symbol interference (ISI) through zero-forcing (ZF) equalization. Focusing on Gaussian transmit pulse shaping filters, we derive an approximate analytical expression for the symbol error probability (SEP) of the proposed receiver and validate its accuracy through Monte Carlo simulations. Our analysis further reveals a trade-off between transmit pulse shaping bandwidth and detection robustness, highlighting important considerations in the design of event-driven receivers for practical communication systems.

\emph{Notation:} The following notations are used throughout the paper. Vectors and matrices are written in bold format (e.g., $\mathbf{z}$, $\mathbf{A}$) while sets are written in calligraphic format (e.g., $\mathcal{S}$). The inverse of a square matrix $\mathbf{A}$ is written as $\mathbf{A}^{-1}$ while its transpose is written as $\mathbf{A}^{T}$. The $l$-th diagonal element of a square matrix $\mathbf{A}$ is denoted as $[\mathbf{A}]_{l,l}$. We use the notation $\mathcal{Q}(\cdot)$ to refer to the Q-function, i.e., the tail probability of the standard Gaussian distribution.

 \begin{figure*}[ht!]
    \centering
    \includegraphics[scale = .8]{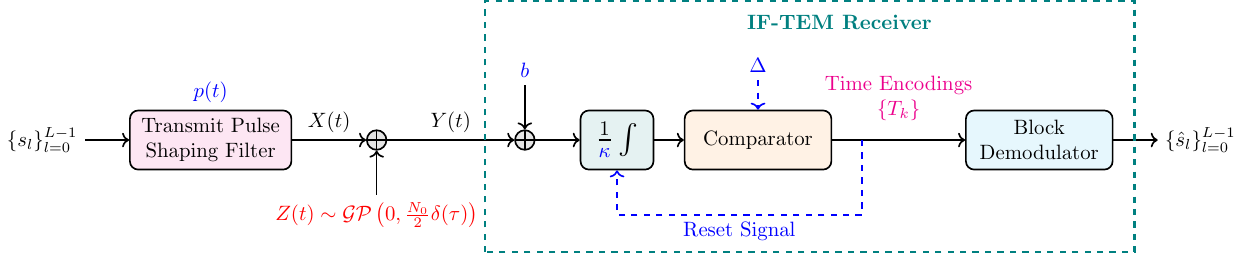}
    \caption{Communication model for the proposed IF-TEM receiver.}
    \label{fig:IFTEM_model}
\end{figure*}

%% file: Sections/problem_setup.tex
\subsection{Communication Model}

We consider the length-$L$ block-based communication model illustrated in Figure \ref{fig:IFTEM_model}. The $L$ symbols $\{s_l\}_{l = 0}^{L-1}$ are drawn independently from an $M$-ary equidistant PAM constellation set $\mathcal{A} = \{a_1,\cdots,a_{M}\}$ with zero mean and unitary average symbol energy. This length-$L$ symbol sequence is then passed through a transmit pulse shaping filter $p(t)$, resulting in the continuous-time signal
\begin{align}\label{eq:tx_sig_model}
    X(t) = \sum_{l = 0}^{L-1}s_l\cdot p(t - l\cdot T_{\mathrm{sym}}),
\end{align}
where $T_{\mathrm{sym}}$ is the symbol period. In this work, we employ a Gaussian pulse-shaping filter with an impulse response given by
\begin{align}
    p_{\mathrm{Gauss}}(t) = \frac{\sqrt{\pi}}{a}\exp\left(-\frac{\pi^2t^2}{a^2}\right),
\end{align}
where $a$ is a shaping parameter inversely proportional to the product of the 3 dB bandwidth, denoted $B_{3\mathrm{dB}}$, and symbol period $T_{\mathrm{sym}}$. Specifically, the shaping parameter $a$ is defined as
\begin{align}
    a = \frac{1}{B_{3\mathrm{dB}}\cdot T_{\mathrm{sym}}}\sqrt{\frac{\ln2}{2}}.
\end{align}

The transmitter output $X(t)$ is sent over an additive noise channel, where the additive noise $Z(t)$ is modeled as a zero-mean white Gaussian process with autocorrelation function $R_{Z}(\tau) = \frac{N_0}{2}\delta(\tau)$
and $N_0$ is the noise spectral density (in Watts/Hz). The received signal $Y(t) = X(t) + Z(t)$ is given by the superposition of the transmitted signal and additive noise.

At the receiver, the continuous-time signal \( Y(t) \) is sampled using an IF-TEM sampler with parameters $b$, $\Delta$, and $\kappa$. First, a constant bias \( b \) is added to the received signal, where \(b\) is chosen such that \(\mathbb{P}\left( Y(t) + b < 0\right)\) is sufficiently small. The resulting biased signal is then scaled by \( \kappa \) and fed into an integrator, as illustrated in Figure~\ref{fig:IFTEM_model}. The output of the integrator is continuously compared against a threshold \( \Delta \). Each time the integrator output reaches the threshold \(\Delta\), the corresponding time instant \( \{t_k\} \) (also referred to as \emph{firing instance}) is recorded. A reset signal for the integrator is triggered after each firing. The firing instances $\{t_k\}$ satisfy the following equilibrium condition for all $k$:
\begin{align}\label{eq:equilibrium_cond}
    \Delta = \frac{1}{\kappa}\int_{t_{k-1}}^{t_k}\left[Y(t) + b\right]dt.
\end{align}
Due to the unbounded nature of the firing instances, it is often preferable to record the time difference between consecutive firings, called \emph{time encodings}. It is known that the time encodings $T_{k} = t_{k} - t_{k-1}$ are bounded by
\begin{align*}
    \frac{\kappa\Delta}{b+c_{\max}} \leq T_{k} \leq \frac{\kappa\Delta}{b-c_{\max}},
\end{align*}
for all $k$, where $c_{\max}$ is the maximum amplitude of the IF-TEM input \cite{Lazar:2004}. The block demodulator estimates the length-$L$ symbol sequence sent by the transmitter using the time encodings $\{T_k\}$ produced by the IF-TEM sampler.

 \subsection{Symbol Sequence Recovery from Time Encodings}

 We present a mechanism to recover the length-$L$ symbol sequence directly from the time encodings $\{T_k\}$. Since iterative signal reconstruction from time encodings \cite{Lazar:2004} is computationally expensive, we avoid this reconstruction step and show that signal reconstruction from time encodings is not necessary for the symbol detection task. The core idea behind the proposed symbol recovery scheme is that the amplitude information carried by \( Y(t) \) over a symbol period is approximately proportional to the number of firing instances occurring within that period. 
 
 Given the time encodings $\{T_k\}$, we first recover the firing instances $\{t_{k}\}$ using the recurrence relation
 \begin{align}
     t_{k} = \begin{cases}
         -0.5T_{\mathrm{sym}}\;,\; k = 0\\
         \;t_{k-1} + T_{k},\;\mathrm{otherwise}
     \end{cases}.
 \end{align}
 Here, we assume that the IF-TEM starts to record events at $t_0 = -0.5T_{\mathrm{sym}}$. Thus, the integrator is at rest at $t = t_0$. Estimation of the symbol timing offset at the receiver using the time encodings will be explored in our future work.
 
 Let $N^{(l)}$ denote the number of firing instances within the interval $\mathcal{T}_l = [(l-0.5)T_{\mathrm{sym}},(l+0.5)\cdot T_{\mathrm{sym}})$ which is centered around the $l$-th symbol. Due to the equilibrium condition in \eqref{eq:equilibrium_cond}, the following relationship holds:
 \begin{align}\label{eq:symbol_recovery_cond}
    N^{(l)}\Delta\kappa &= \int_{t_{\mathrm{min}}^{(l)}}^{t_{\mathrm{max}}^{(l)}}\left[X(t) + Z(t)\right]dt
    +b\left(t_{\mathrm{max}}^{(l)}-t_{\mathrm{min}}^{(l)}\right)
\end{align}
where $t_{\mathrm{min}}^{(l)} = \underset{t_{k}\in\mathcal{T}_l}{\min}\; t_k$ and $t_{\mathrm{max}}^{(l)} = \underset{t_{k}\in\mathcal{T}_l}{\max}\; t_k$. Since $Z(t)$ is a zero-mean white Gaussian process, the time average of $Z(t)$ from $t_{\mathrm{min}}^{(l)}$ to $t_{\mathrm{max}}^{(l)}$ follows a zero-mean Gaussian distribution. More precisely, we have
\begin{align*}
    Z'_l = \frac{\int_{t_{\mathrm{min}}^{(l)}}^{t_{\mathrm{max}}^{(l)}}Z(t)dt}{t_{\mathrm{max}}^{(l)}-t_{\mathrm{min}}^{(l)}} \;\sim\; \mathcal{N}\left(0,\frac{N_0}{2[t_{\mathrm{max}}^{(l)}-t_{\mathrm{min}}^{(l)}]}\right).
\end{align*}
Dividing both sides of \eqref{eq:symbol_recovery_cond} by $t_{\mathrm{max}}^{(l)}-t_{\mathrm{min}}^{(l)}$, and then subtracting the bias term from both sides give us
\begin{align}\label{eq:symbol_recovery_cond2}
    \frac{N^{(l)}\Delta\kappa}{t_{\mathrm{max}}^{(l)}-t_{\mathrm{min}}^{(l)}} - b &= \frac{1}{t_{\mathrm{max}}^{(l)}-t_{\mathrm{min}}^{(l)}}\int_{t_{\mathrm{min}}^{(l)}}^{t_{\mathrm{max}}^{(l)}}X(t)dt
    +Z'_l.
\end{align}

Note that the objective is to recover the length-$L$ symbol sequence $\{s_{l}\}_{l = 0}^{L-1}$ embedded in $X(t)$. Suppose we define a $L\times L$ matrix $\mathbf{P}$ whose $(l+1)$-th row and $(j+1)$-th column is given by
\begingroup
\allowdisplaybreaks
\begin{align}\label{eq:P_ij}
    P_{l,j} =& \frac{1}{t_{\mathrm{max}}^{(j)}-t_{\mathrm{min}}^{(j)}}\int_{t_{\mathrm{min}}^{(j)}}^{t_{\mathrm{max}}^{(j)}}p(t - l\cdot  T_{\mathrm{sym}})dt\nonumber\\
    =& \frac{\mathcal{Q}\left(\frac{ t_{\mathrm{min}}^{(j)} - l\cdot T_{\mathrm{sym}}}{a/\sqrt{2}\pi}\right) - \mathcal{Q}\left(\frac{ t_{\mathrm{max}}^{(j)} - l\cdot T_{\mathrm{sym}}}{a/\sqrt{2}\pi}\right)}{\left(t_{\mathrm{max}}^{(j)}-t_{\mathrm{min}}^{(j)}\right)}.
\end{align}
\endgroup
 The second line follows since we used a Gaussian pulse shaping filter. By plugging in \eqref{eq:tx_sig_model} to \eqref{eq:symbol_recovery_cond2}, we obtain
\begin{align}\label{eq:symbol_recovery_cond3}
    \frac{N^{(l)}\Delta\kappa}{t_{\mathrm{max}}^{(l)}-t_{\mathrm{min}}^{(l)}} - b &= \sum_{j = 0}^{L-1}s_{j}\cdot\frac{\int_{t_{\mathrm{min}}^{(j)}}^{t_{\mathrm{max}}^{(j)}}p(t-l\cdot T_{\mathrm{sym}})dt}{t_{\mathrm{max}}^{(j)}-t_{\mathrm{min}}^{(j)}}
     +Z'_l\nonumber\\
     &= \langle\mathbf{p}_l,\mathbf{s}\rangle + Z_l',
\end{align}
where the second line comes from the definition of the inner product. The vector $\mathbf{s} = [s_0,s_1,\cdots,s_{L-1}]^T$ is the vectorized form of the symbol sequence $\{s_{l}\}_{l = 0}^{L-1}$ and $\mathbf{p}_l = [P_{l,0},P_{l,1},\cdots,P_{l,L-1}]^T$ contains the $(l+1)$-th row of $\mathbf{P}$. Let $\mathbf{Z}' = [Z_0',Z_1',\cdots,Z_{L-1}']^T$ be the vectorized form of $\{Z_l'\}_{l = 0}^{L-1}$ and let $\mathbf{y} = [y_0,y_1,\cdots,y_{L-1}]^T$ be a $L\times 1$ vector whose $l$-th element is
\begin{align*}
    y_l = \frac{N^{(l)}\Delta\kappa}{t_{\mathrm{max}}^{(l)}-t_{\mathrm{min}}^{(l)}}.
\end{align*}
Then, \eqref{eq:symbol_recovery_cond3} can be represented in matrix form as follows:
\begin{align}\label{eq:GLM}
    \mathbf{y} = \mathbf{P}\mathbf{s} + \mathbf{Z}'.
\end{align}
The above matrix representation resembles a Gaussian linear model (GLM) and a non-diagonal matrix $\mathbf{P}$ means that ISI is present in the observation vector $\mathbf{y}$. Various linear and nonlinear detection strategies can be used to recover $\mathbf{s}$ from $\mathbf{y}$. To facilitate recovery, we adopt a simple ZF detector. Specifically, we post-process the observation vector \( \mathbf{y} \) to obtain the ZF estimate: 
\begin{align*}
    \mathbf{\tilde{s}}_{\mathrm{ZF}} = \left(\mathbf{P}^T \mathbf{P} \right)^{-1} \mathbf{P}^T \mathbf{y},
\end{align*}
Finally, each element of \( \mathbf{\tilde{s}}_{\mathrm{ZF}} \) is passed through an \( M \)-PAM hard decision decoder to produce the length-\( L \) symbol sequence estimate \( \{\hat{s}_l\}_{l = 0}^{L-1} \).

%% file: Sections/error_analysis.tex
In this section, we analyze the SEP of the communication model described in Section \ref{section:sys_model}. The exact SEP is difficult to analyze because of the dependence of $\mathrm{Var}\left(Z_{l}'\right)$ and $\mathbf{P}$ on $t_{\mathrm{max}}^{(l)}-t_{\mathrm{min}}^{(l)}$. As an alternative, we use the approximations $t_{\mathrm{max}}^{(l)} \approx (l+0.5)T_{\mathrm{sym}}$ and $t_{\mathrm{min}}^{(l)} \approx (l-0.5)T_{\mathrm{sym}}$ for all $l$. These approximations are tight if the ratio $\frac{N^{(l)}}{T_{\mathrm{sym}}}$ is large for all $l$. Consequently, $Z'_l\sim\mathcal{N}\left(0,\frac{N_0}{2T_{\mathrm{sym}}}\right)$ and the entry in the $(l+1)$-th row and $(j+1)$-th column of $\mathbf{P}$ becomes
\begin{align}\label{eq:P_ij_approx}
    P_{l,j} =& \frac{\mathcal{Q}\left(\frac{  (j-l - 0.5)T_{\mathrm{sym}}}{a/\sqrt{2}\pi}\right) - \mathcal{Q}\left(\frac{  (j-l+0.5)T_{\mathrm{sym}}}{a/\sqrt{2}\pi}\right)}{T_{\mathrm{sym}}}.
\end{align}

In this paper, we define the probability of symbol error as
 \begin{align}\label{eq:general_Pe}
    P_{\mathrm{e}} =& \frac{1}{L}\sum_{l = 0}^{L-1}\mathbb{P}\left(\hat{S}_{l} \neq S_{l}\right).
\end{align}
% \begin{align*}
%     \mathbf{\tilde{S}}_{\mathrm{ZF}} = \mathbf{S} + \mathbf{\tilde{Z}}',
% \end{align*}
Due to the ZF solution, the ISI is removed at the expense of potentially introducing correlation among noise components $\mathrm{Z}_l'$. More precisely, the ZF estimate can be written as
$\mathbf{\tilde{S}}_{\mathrm{ZF}} = \mathbf{S} + \mathbf{\tilde{Z}}'$,
where $\mathbf{\tilde{Z}}'\sim\mathcal{N}\left(\mathbf{0},\frac{N_0}{2T_{\mathrm{sym}}}\left(\mathbf{P}^T\mathbf{P}\right)^{-1}\right)$ is the resulting noise vector after the ZF equalization. Let $\tilde{S}_{\mathrm{ZF}}^{(l)}$ be the $l$-th element of $\mathbf{\tilde{S}}_{\mathrm{ZF}}$ and $\boldsymbol{\Sigma}_{\tilde{Z}'} = \frac{N_0}{2T_{\mathrm{sym}}}\left(\mathbf{P}^T\mathbf{P}\right)^{-1}$ be the covariance matrix of $\mathbf{\tilde{Z}}'$. After applying the ZF equalizer, the post-equalization signal-to-noise ratio (SNR) at the $l$-th symbol period becomes
\begin{align}
    \gamma_{l} = \frac{1}{\left[\boldsymbol{\Sigma}_{\tilde{Z}'}\right]_{l,l}} = \frac{2T_{\mathrm{sym}}}{N_0}\left[\left(\mathbf{P}^T\mathbf{P}\right)\right]_{l,l} .
\end{align}
Since $S_{l}$ is drawn from a unit-energy $M$-PAM constellation set $\mathcal{A}$, the probability that the hard decision decoder output $\hat{S}_{l}$ is not equal to $\mathrm{S}_{l}$ can be expressed as \cite{Proakis2007}:
\begin{align}\label{eq:PAM_SEP}
    \mathbb{P}\left(\hat{S}_{l} \neq S_{l}\right) = \frac{2(M-1)}{M}\mathcal{Q}\left(\sqrt{\frac{3\gamma_{l}}{M^2-1}}\right)
\end{align}
Observe that the dependence of $\mathbb{P}\left(\hat{S}_{l} \neq S_{l}\right)$ to $l$ is through the post-equalization SNR $\gamma_{l}$. Combining equations \eqref{eq:general_Pe} and \eqref{eq:PAM_SEP} gives us
\begin{align}\label{eq:TEM_SEP}
    P_{\mathrm{e}} =& \frac{2(M-1)}{M\cdot L}\sum_{l = 0}^{L-1}\mathcal{Q}\left(\sqrt{\frac{3\gamma_{l}}{M^2-1}}\right).
\end{align}
It is important to note that the impact of the pulse shaping filter $p(t)$ on $P_{\mathrm{e}}$ is through the post-equalization SNR values $\{\gamma_{l}\}_{l = 0}^{L-1}$. This is investigated in the next section.

%% file: Sections/numerical_result.tex
\begin{table}[t]
\centering
\caption{Condition number $\kappa_{\mathrm{cn}}\left(\mathbf{P}\right)$ for different settings of $B_{\mathrm{3dB}}$ and $L$}
\begin{tabular}{|c|c|c|c|c|c|c|}
\hline
\multirow{2}{*}{} & \multicolumn{6}{c|}{\textbf{3 dB Bandwidth} (in Hz)} \\
\cline{2-7}
& 0.125 & $0.25$ & $0.50$ & $0.75$ & $1.00$ & $1.25$ \\
\hline
$L = 10$ & 110.4154 & 2.9600 & 1.1258 & 1.0086 & 1.0003 & 1.0000 \\
\hline
$L = 50$ & 194.1509 & 3.1168 & 1.1312 & 1.0090 & 1.0003 & 1.0000 \\
\hline
$L = 100$ & 198.5801 & 3.1240 & 1.1314 & 1.0090 & 1.0003 & 1.0000 \\
\hline
\end{tabular}
\label{tab:condP-vs-bandwidth}
\end{table}

In this section, we validate the theoretical SEP derived for the proposed IF-TEM receiver. Numerical experiments are also conducted to gain additional insight on how the pulse shaping filter affects the error probability of the system.

We first analyze the noise sensitivity of the Gaussian pulse for different settings of the shaping parameter. The symbol period is set to $T_{\mathrm{sym}} = 1$ second/symbol so that the shaping parameter solely depends on the 3 dB bandwidth $B_{3\mathrm{dB}}$. We use \eqref{eq:P_ij_approx} to approximate the entries of $\mathbf{P}$ and we set $t_{\mathrm{min}}^{(l)} = (l-0.5)T_{\mathrm{sym}}$ and $t_{\mathrm{max}}^{(l)} = (l+0.5)T_{\mathrm{sym}}$. The noise sensitivity of the Gaussian pulse is measured through the condition number of $\mathbf{P}$, i.e. $\kappa_{cn}\left(\mathbf{P}\right)$. High condition number means that $\mathbf{P}$ is ill-conditioned and applying ZF equalization would significantly amplify the noise. In contrast, a condition number close to unity would imply that pulse shape is robust against noise. 

Table \ref{tab:condP-vs-bandwidth} lists the condition number $\kappa_{cn}\left(\mathbf{P}\right)$ for different settings of $B_{\mathrm{3dB}}$ and $L$. It can be observed that reducing the 3 dB bandwidth of the Gaussian pulse makes the system more sensitive to noise. This can be attributed to the significant ISI in Gaussian pulses with large shaping parameter $a$. At $B_{\mathrm{3dB}} = 0.125$, it can be seen that the matrix $\mathbf{P}$ is ill-conditioned. We also observed that increasing the block length $L$ slightly increases the condition number of $\mathbf{P}$. However, the increase is not significant and the condition number is less affected by $L$ for large values of the 3 dB bandwidths.

Next, we validate the theoretical SEP result in \eqref{eq:TEM_SEP} by comparing it with the simulated performance of the proposed IF-TEM receiver. We choose $L = 50$ and $M = 4$ for the block-based PAM transmission. The symbol period is fixed to $T_{\mathrm{sym}} = 1$ and we consider three different values of the 3 dB bandwidth: $B_{\mathrm{3 dB}} = 0.25, 0.50,\;\mathrm{and}\;1.00$. Since $E_{\mathrm{s}} = 1$, the per-symbol SNR, denoted $E_{\mathrm{s}}/N_0$, is varied by changing the noise spectral density. For the IF-TEM implementation, the time encoding and decoding toolkit \cite{bionet_ted_matlab} from the Bionet Group of Columbia University  is used. The IF-TEM parameters are set to $\Delta = 1$, $b = 5$, and $\kappa = 0.01$. The choice of the IF-TEM bias parameter ensures that \(\mathbb{P}\left( Y(t) + b < 0\right)\) is sufficiently small, i.e.,  $\ll 1\times 10^{-3}$.

\begin{figure}[t!]
    \hspace{-.425cm}
    \includegraphics[scale = .51]{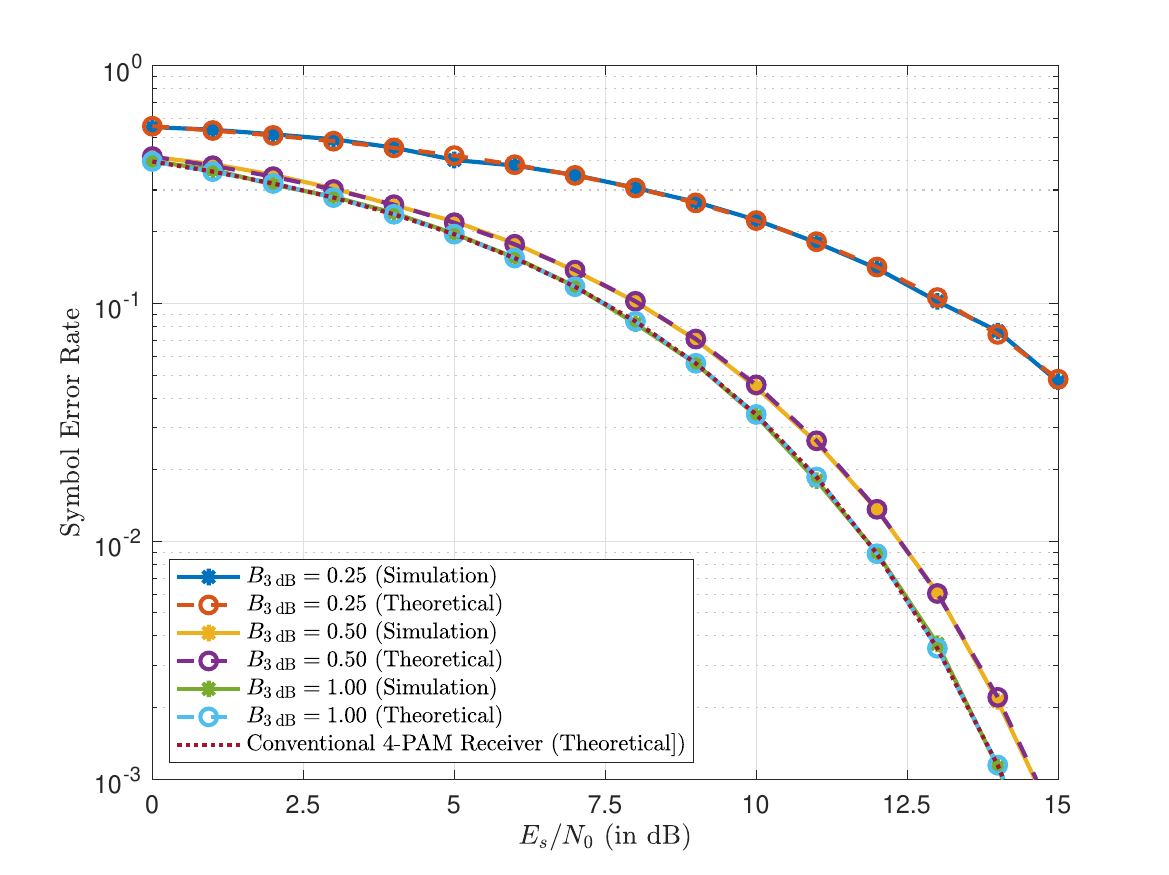}
    \caption{Symbol Error Probability vs. $E_{\mathrm{s}}/N_0$ (in dB) of the proposed IF-TEM receiver under different 3 dB bandwidth settings ($B_{\mathrm{3dB}} = 0.25, 0.50, 1.00$).}
    \label{fig:SEP_curves}
\end{figure}

Figure \ref{fig:SEP_curves} depicts the symbol error probability curves of the proposed IF-TEM receiver for 4-PAM under different settings of the 3 dB bandwidth. Despite the approximations for $t_{\mathrm{max}}^{(l)}$ and $t_{\mathrm{min}}^{(l)}$ used in Section \ref{section:error_analysis}, it can be observed that the simulated performance of the proposed IF-TEM receiver is close to the theoretical prediction established in equation \eqref{eq:TEM_SEP}. The SEP curve for $B_{\mathrm{3dB}} = 1.00$ coincides with the theoretical SEP of a conventional 4-PAM receiver. Moreover, the SEP values increase as we reduce the 3 dB bandwidth. This performance degradation can be attributed to $\kappa_{cn}(\mathbf{P})$.

%the increase in $\kappa_{cn}(\mathbf{P})$.

%% file: Sections/conclusion.tex
In this work, we proposed an IF-TEM-based PAM receiver that estimates a length-$L$ sequence of PAM symbols directly from the time encodings $\{T_k\}$. The proposed receiver uses the number of firing instances within each symbol period as a decision statistic for symbol detection. The ISI introduced by the transmit pulse shaping filter is mitigated using ZF equalization. We derived an approximate expression for the SEP of the IF-TEM receiver and showed, through numerical simulations, that the analytical approximation closely matches the empirical SEP. Furthermore, we observed that narrowing the 3 dB bandwidth of the transmit pulse shaping filter decreases the system’s resilience to noise, highlighting a trade-off between spectral efficiency and robustness. These findings underscore the importance of jointly optimizing pulse shaping, equalization, and time encoding in the design of time-based receivers. Overall, our results demonstrate the feasibility of doing symbol sequence detection directly from time encodings, laying the groundwork for low-power, event-driven receiver architectures.

Future work will explore more sophisticated decision strategies that leverage the full temporal structure of the time encodings beyond simple event counting, such as maximum likelihood sequence decoding. Additionally, the performance of IF-TEM receivers under more realistic channel conditions (e.g. multipath fading) and hardware model (e.g. carrier and symbol timing offset) is an open research topic. Finally, implementing the IF-TEM receiver on hardware platforms, such as field-programmable gate arrays (FPGAs) or neuromorphic processors, will be crucial to assess its practical viability in low-power communication systems.